\documentclass[preprint]{elsarticle}
\usepackage{graphicx}
\usepackage{amsmath,amssymb}
\DeclareMathOperator{\tr}{Tr}

\begin{document}
\begin{frontmatter}
\title{Decoherence of Coupled Quantum Oscillators: Going Beyond the Caldeira-Leggett Approximation}
\author{Bj\"{o}rn Bartels}
\ead{bbartels@physik.uni-wuerzburg.de}
\address{Institut f\"{u}r Theoretische Physik, Universit\"{a}t W\"{u}rzburg, D-97074 W\"{u}rzburg, Germany}

\begin{abstract}

We calculate the reduced density matrix for a system of coupled harmonic oscillators in a bosonic heat bath using the Born-Markov approximation and show that the expectation values of position and momentum evolve like classical quantities. We consider the cases where every oscillator is coupled to its own reservoir and where all oscillators are coupled to the same heat bath, pointing out the differences between these two models. In particular, for systems coupled to a common reservoir, we present an uncommon dissipation mechanism, which disappears, when the dynamics of the system are gouverned by the low frequency modes of the environment (Caldeira-Leggett limit). We also address the interesting phenomenon of reservoir-induced interactions.

\end{abstract}

\begin{keyword}
decoherence\sep Quantum Brownian Motion\sep Born-Markov approximation\sep interacting quantum systems
\end{keyword}

\end{frontmatter}

\section{Introduction}

In the early 1980s decoherence theory made a first attempt to explain the emergence of classicality in quantum mechanics \cite{zu1}. While the quantum-to-classical transition is today well understood \cite{sc1,sc2,joo,zu2}, decoherence is nowadays the major obstacle in the experimental implementation of quantum information processing \cite{nie}. It is therefore a challenging task to find mechanisms that suppress decoherence. Recently, {\it dissipation and decoherence free subspaces} have been found in oscillator networks \cite{po1}. The underlying mechanism is that in multipartite systems interactions between the subsystems can lead to the robustness of certain states, while other states remain affected by decoherence. A second reason why one is interested in the decoherence of interacting quantum systems is {\it quantum synchronization}. In classical mechanics, two coupled dissipative systems can evolve into a synchronized state \cite{bal} and will stay there for all times. If one wants to find a quantum analogue of synchronization, one is confronted with the problem that quantum mechanics is invariant by time reversal, which provides no reason why two initially synchronized {\it quantum} systems could not desynchronize again, simply by reversing the arrow of time. It is suggested that decoherence, i. e. the interaction of a system with its environment, as it breaks time reversal symmetry, is an essential ingredient in order to study quantum synchronization \cite{zhi,ort}.\\
\indent As the interaction of a system with its environment can be very complicated, it is clear that analytical calculations are limited to a very small class of models. One of the simplest models is a harmonic oscillator linearly coupled to a bosonic heat bath and has been studied in great detail in literature under the name of {\it Quantum Brownian Motion} \cite{cal,gra,unr,hu,wei}. Although some work has already been done on the decoherence of interacting quantum systems \cite{zha,cho,po2}, the problem of several interacting harmonic oscillators is not yet fully understood. As an important contribution, Cacheffo et al. \cite{cac} solved the problem in the so-called Caldeira-Leggett limit. While these authors use path integral methods to derive the reduced density matrix \cite{fey}, in the present work we will apply the Born-Markov approximation, which not only reproduces the results of \cite{cac} but also gives more general results and is less complicated from a mathematical point of view. In the present paper, we will derive and solve the Born-Markov equation of $N$ linearly coupled harmonic oscillators, each one interacting with its own bosonic heat bath, and point out the formal analogy with the problem of a single oscillator in a heat bath (section \ref{s2}). After that, we will compare the result for distinct reservoirs with that obtained for the coupling to a common reservoir (section \ref{s3}). In detail, we will consider a general spectral density and analyze the behaviour of the expectation values of position for the case of two interacting oscillators (section \ref{s4}).

\section{Distinct Reservoirs}\label{s2}

In the present section we consider the system Hamiltonian
\begin{equation}
H_\mathcal{S}=\sum_{\alpha=1}^N\left(\frac{P_\alpha^2}{2M_\alpha} +\frac{1}{2}M_\alpha\Omega_\alpha^2 X_\alpha^2\right)+\frac{1}{2}\sum_{\alpha\neq\beta}g_{\alpha\beta}X_\alpha X_\beta,
\end{equation}
where $X_\alpha$ and $P_\alpha$ are the position and momentum operators of $N$ harmonic oscillators with frequencies $\Omega_\alpha$ and masses $M_\alpha$, linearly coupled by coupling constants $g_{\alpha\beta}=g_{\beta\alpha}$. Every oscillator $\alpha$ is coupled to its own bosonic heat bath
\begin{equation}H_\mathcal{E}^{(\alpha)}=\sum_{j=1}^{N_{HB}^{(\alpha)}}\left(\frac{p_j^{(\alpha)^2}}{2m_j^{(\alpha)}}+\frac{1}{2}m_j^{(\alpha)} \omega^{(\alpha)^2}_j x^{(\alpha)^2}_j\right)
\end{equation}
(where $\omega_j^{(\alpha)}$ are the frequencies and $m_j^{(\alpha)}$ the masses of $N_{HB}^{(\alpha)}$ non-interacting harmonic oscillators) via an interaction Hamiltonian
\begin{equation}
 H_I^{(\alpha)}= X_\alpha\otimes\sum_{j=1}^{N_{HB}^{(\alpha)}} c^{(\alpha)}_j x^{(\alpha)}_j\equiv X_\alpha\otimes E_\alpha
\end{equation}
with coupling constants $c_j^{(\alpha)}$. The total Hamiltonian is therefore:
\begin{equation}
 \label{ham}H=H_\mathcal{S}+H_\mathcal{E}+H_I
\end{equation}
with
\begin{equation}
 H_\mathcal{E}=\sum_{\alpha=1}^N H_\mathcal{E}^{(\alpha)},\qquad  H_I=\sum_{\alpha=1}^N H_I^{(\alpha)}.
\end{equation}

\subsection{Derivation of Born-Markov Equation}

In order to determine the temporal evolution of the system described by \eqref{ham}, it is useful to have a master equation for the reduced density matrix $\rho_\mathcal{S}$ of the system. In the case of Quantum Brownian Motion an exact master equation has been derived by Hu, Paz and Zhang \cite{hu}. A much simpler form can be obtained by applying the Born-Markov approximation, which means that the system-environment interaction is weak enough so that the full density matrix $\rho$ of the system {\it and} the environment approximately factorizes for all times (Born approximation):
\begin{equation}\label{bor}
\rho(t)\approx\rho_\mathcal{S}(t)\otimes\rho_\mathcal{E}(0) 
\end{equation}
(where $\rho_\mathcal{E}(0)$ is the initial state of the environment) and that the correlations $\langle E_\alpha^{(I)}(t)E_\beta^{(I)}(t')\rangle_{\rho_\mathcal{E}}$
of the environment decay more rapidly than $\rho_\mathcal{S}$ changes (Markov approximation). Here, $\langle\cdot\rangle_{\rho_\mathcal{E}}$ denotes the expectation value with respect to $\rho_\mathcal{E}(0)$, while the index $(I)$ denotes the temporal evolution in the interaction picture, i. e.
\begin{equation}
E_\alpha^{(I)}(\tau) =\mathrm e^{i (H_\mathcal{S}+H_\mathcal{E})\tau}E_\alpha e^{-i (H_\mathcal{S}+H_\mathcal{E})\tau}.
\end{equation}
Using the Born-Markov approximation, one can derive the following master equation \cite{sc1}:
\begin{equation}\label{bmp}\begin{split}\frac{\mathrm d}{\mathrm d t}\rho_\mathcal{S}(t)=-i\left[H_\mathcal{S},\rho_\mathcal{S}(t)\right]-\sum_{\alpha,\beta}\int_0^\infty\mathrm d \tau & \left(\mathcal{C}_{\alpha\beta}(\tau)[X_\alpha,X_\beta^{(I)}(-\tau)\rho_\mathcal{S}(t)]\right.\\
& \left.+\mathcal{C}_{\alpha\beta}(-\tau)[\rho_\mathcal{S}(t)X_\beta^{(I)}(-\tau),X_\alpha]\right),\end{split}\end{equation}
where 
\begin{equation}
 \mathcal{C}_{\alpha\beta}(\tau):=\left\langle E^{(I)}_\alpha(\tau)E_\beta^{(I)}(0)\right\rangle_{\rho_\mathcal{E}}
\end{equation}
is the correlation function between the baths at time difference $\tau$.\\
\indent Expressing $x_j^{(\alpha)}$ and $p_j^{(\alpha)}$ in terms of annihilation and creation operators ($x_j^{(\alpha)}=(b_j^{(\alpha)}+b_j^{(\alpha)^\dagger})/\sqrt{2m_j^{(\alpha)}\omega_j^{(\alpha)}}$, $p_j^{(\alpha)}=i\sqrt{m_j^{(\alpha)}\omega_j^{(\alpha)}/2}(b_j^{(\alpha)^\dagger}-b_j^{(\alpha)})$) and using the Baker-Campbell-Hausdorff formula
\begin{equation}
 \label{bch}e^X Y e^{-X}=\sum_{n=0}^\infty \frac{1}{n!}[X,Y]_n
\end{equation}
(with the nested commutator $[X,Y]_n:=[X,[X,Y]_{n-1}]$, $[X,Y]_0:=Y$), one finds for the correlation function
\begin{equation}
 \begin{split}\mathcal{C}_{\alpha\beta}(\tau)=\sum_{j=1}^{N_{HB}^{(\alpha)}}\sum_{l=1}^{N_{HB}^{(\beta)}}\frac{c_j^{(\alpha)} c_l^{(\beta)}}{2\sqrt{m_j^{(\alpha)}\omega_j^{(\alpha)}m_l^{(\beta)}\omega_l^{(\beta)}}} & \left\langle\left(b_j^{(\alpha)^\dagger}e^{i\omega_j^{(\alpha)}\tau}+b_j^{(\alpha)}e^{-i\omega_j^{(\alpha)}\tau}\right)\right.\\
& \qquad\qquad\qquad\times\left.\left(b_l^{(\beta)}+b_l^{(\beta)^\dagger}\right)\right\rangle_{\rho_\mathcal{E}}.\end{split}
\end{equation}
Inserting the equilibrium density matrix
\begin{equation}
 \rho_\mathcal{E}=\prod_{\alpha=1}^N \frac{e^{-H_\mathcal{E}^{(\alpha)}/T_\alpha}}{\tr_{\mathcal{E}_\alpha}e^{-H_\mathcal{E}^{(\alpha)}/T_\alpha}}
\end{equation}
(where $T_\alpha$ is the temperature of the $\alpha$-th heat bath, $k_B\equiv 1$) and evaluating the trace in the occupation number representation, one immediately sees that
\begin{equation}
 \left\langle b_j^{(\alpha)}b_l^{(\beta)}\right\rangle_{\rho_\mathcal{E}}=\left\langle b_j^{(\alpha)^\dagger}b_l^{(\beta)^\dagger}\right\rangle_{\rho_\mathcal{E}}=0\end{equation}
\begin{equation}
 \left\langle b_j^{(\alpha)^\dagger}b_l^{(\beta)}\right\rangle_{\rho_\mathcal{E}}=\frac{\delta_{\alpha\beta}\delta_{jl}}{e^{\omega_j^{(\alpha)}/T_\alpha}-1}\qquad \left\langle b_j^{(\alpha)}b_l^{(\beta)^\dagger}\right\rangle_{\rho_\mathcal{E}}=\delta_{\alpha\beta}\delta_{jl}\left(1+\frac{1}{e^{\omega_j^{(\alpha)}/T_\alpha}-1}\right)
\end{equation}
($j=1,\dots,N_{HB}^{(\alpha)}$ and $l=1,\dots,N_{HB}^{(\beta)}$).
The correlation function can then be written as
\begin{equation}
 \label{cod}\mathcal{C}_{\alpha\beta}(\tau)=\delta_{\alpha\beta}(\nu_\alpha(\tau)-i\eta_\alpha(\tau))
\end{equation}
with the \emph{noise kernel}
\begin{equation}\label{nk}\nu_\alpha(\tau):= \sum_{j=1}^{N_{HB}^{(\alpha)}} \frac{c_j^{(\alpha)^2}}{2m_j^{(\alpha)}\omega_j^{(\alpha)}}\coth\left(\frac{\omega_j^{(\alpha)}}{2T_\alpha}\right)\cos(\omega_j^{(\alpha)}\tau)
 \end{equation}
and the \emph{dissipation kernel}
\begin{equation}
 \label{dk}\eta_\alpha(\tau):= \sum_{j=1}^{N_{HB}^{(\alpha)}} \frac{c_j^{(\alpha)^2}}{2m_j^{(\alpha)}\omega_j^{(\alpha)}}\sin(\omega_j^{(\alpha)}\tau).
\end{equation}
\indent Calculating $X_\alpha^{(I)}(\tau)$ is a bit more complicated. Again it is  easier to work with annihilation and creation operators $a_\alpha$, $a_\alpha^\dagger$ ($\alpha=1,\dots,N$). Observing that
\begin{equation}
 [H_\mathcal{S},a_\alpha]=-\Omega_\alpha a_\alpha-\sum_{\beta=1,\,\beta\neq\alpha}^N\tilde{g}_{\alpha\beta}(a_\beta+a_\beta^\dagger)
\end{equation}
(with $\tilde{g}_{\alpha\beta}=g_{\alpha\beta}/\sqrt{4M_\alpha\Omega_\alpha M_\beta\Omega_\beta}$), one is led to
\begin{equation}
 \label{it}[H_\mathcal{S},a_\alpha]_n=\mathbf{h}_n^{(\alpha)}\cdot\mathbf{a},
\end{equation}
where $\mathbf{a}=(a_1^\dagger,a_1,\dots,a_N^\dagger,a_N)^T$ and $\mathbf{h}_n^{(\alpha)}$ is a vector valued function depending on $\Omega_\alpha$ and $\tilde{g}_{\alpha\beta}$ ($\alpha,\beta=1,\dots,N$). By iterating \eqref{it} one finds that
\begin{equation}
 [H_\mathcal{S},a_\alpha]_n=\mathbf{G}^n\mathbf{h}_0^{(\alpha)}\cdot\mathbf{a},
\end{equation}
where the matrix $\mathbf{G}$ is defined by
\begin{align}
\mathbf{G}:=\begin{pmatrix}\mathbf{W}_1 & \mathbf{\Gamma}_{12} & \cdots & \mathbf{\Gamma}_{1N}\\ \mathbf{\Gamma}_{12} & \ddots & \ddots & \vdots\\ \vdots & \ddots & \ddots & \mathbf{\Gamma}_{N-1,N}\\ \mathbf{\Gamma}_{1N} & \cdots & \mathbf{\Gamma}_{N-1,N} & \mathbf{W}_N\end{pmatrix}\\
\mathbf{W}_\alpha:=\begin{pmatrix}\Omega_\alpha & 0\\ 0 & -\Omega_\alpha\end{pmatrix}\qquad\mathbf{\Gamma}_{\alpha\beta}:=\begin{pmatrix}\tilde{g}_{\alpha\beta} & -\tilde{g}_{\alpha\beta}\\ \tilde{g}_{\alpha\beta} & -\tilde{g}_{\alpha\beta}\end{pmatrix}
\end{align}
and $\left(\mathbf{h}_0^{(\alpha)}\right)_i=\delta_{i,2\alpha}$. It follows that the evolution of $X_\alpha$ is given by
\begin{equation}
\label{xev}\begin{split}X_\alpha^{(I)}(\tau)=\sum_{\beta=1}^N & \left[X_\beta\sqrt{\frac{M_\beta\Omega_\beta}{M_\alpha\Omega_\alpha}}\left(\cos(\mathbf{G}\tau)_{2\beta-1,2\alpha}+\cos(\mathbf{G}\tau)_{2\beta,2\alpha}\right)\right.\\
& \left.+\frac{P_\beta}{\sqrt{M_\alpha\Omega_\alpha M_\beta\Omega_\beta}}\left(\sin(\mathbf{G}\tau)_{2\beta-1,2\alpha}-\sin(\mathbf{G}\tau)_{2\beta,2\alpha}\right)\right],\end{split}
\end{equation}
where $\cos(\mathbf{G}\tau)_{\alpha,\beta}$, $\sin(\mathbf{G}\tau)_{\alpha,\beta}$ are matrix elements of the matrix valued cos/sin function.\\
\indent The Born-Markov master equation \eqref{bmp} can then be written as
\begin{equation}\label{BM}\begin{split}\frac{\mathrm d}{\mathrm d t}\rho_\mathcal{S}(t)=-i[H_\mathcal{S},\rho_\mathcal{S}(t)]-& \sum_{\alpha,\beta=1}^N  \Bigl( D_{\alpha\beta} [X_\alpha,[X_\beta,\rho_\mathcal{S}(t)]]+i\gamma_{\alpha\beta}[X_\alpha,\{P_\beta,\rho_\mathcal{S}(t)\}]\Bigr.\\
& \Bigl.+f_{\alpha\beta} [X_\alpha,[P_\beta,\rho_\mathcal{S}(t)]]+\frac{i}{2}M_\beta\tilde{\Omega}_{\alpha\beta}^2[X_\alpha,\{X_\beta,\rho_\mathcal{S}(t)\}]\Bigr)\end{split}\end{equation}
with the following coefficients:
\begin{align}
\tilde{\Omega}_{\alpha\beta}^2 & :=-\frac{2}{M_\beta}\int_0^\infty\mathrm d\tau\,\eta_{\alpha}(\tau)\sqrt{\frac{M_\beta\Omega_\beta}{M_\alpha\Omega_\alpha}}\left(\cos(\mathbf{G}\tau)_{2\beta-1,2\alpha}+\cos(\mathbf{G}\tau)_{2\beta,2\alpha}\right)\label{coef1}\\
\gamma_{\alpha\beta} & :=\int_0^\infty\mathrm d\tau\,\eta_\alpha(\tau)\frac{1}{\sqrt{M_\alpha\Omega_ \alpha M_\beta\Omega_\beta}}\left(\sin(\mathbf{G}\tau)_{2\beta-1,2\alpha}-\sin(\mathbf{G}\tau)_{2\beta,2\alpha}\right)\\
 D_{\alpha\beta} & :=\int_0^\infty\mathrm d\tau\,\nu_\alpha(\tau)\sqrt{\frac{M_\beta\Omega_\beta}{M_\alpha\Omega_\alpha}}\left(\cos(\mathbf{G}\tau)_{2\beta-1,2\alpha}+\cos(\mathbf{G}\tau)_{2\beta,2\alpha}\right)\\
 f_{\alpha\beta} & :=-\int_0^\infty\mathrm d\tau\,\nu_\alpha(\tau)\frac{1}{\sqrt{M_\alpha\Omega_ \alpha M_\beta\Omega_\beta}}\left(\sin(\mathbf{G}\tau)_{2\beta-1,2\alpha}-\sin(\mathbf{G}\tau)_{2\beta,2\alpha}\right)\label{coef4}
\end{align}

\subsection{Interpretation of Born-Markov Equation}\label{ibm}

Let us now compare the Born-Markov equation \eqref{BM} for coupled oscillators with the one for a single oscillator, which states \cite{sc1}:
\begin{equation}\label{BM1}\begin{split}\frac{\mathrm d}{\mathrm d t}\rho_\mathcal{S}(t)= & -i\left[\frac{P^2}{2M}+\frac{1}{2}M(\Omega^2+\tilde{\Omega}^2) X^2,\rho_\mathcal{S}(t)\right]\\
& -\left( D [X,[X,\rho_\mathcal{S}(t)]]+i\gamma[X,\{P,\rho_\mathcal{S}(t)\}]+f[X,[P,\rho_\mathcal{S}(t)]]\right),\end{split}\end{equation}
where $M$ and $\Omega$ are the mass and frequency of the oscillator and
\begin{align}
\tilde{\Omega}^2 & :=-\frac{2}{M}\int_0^\infty\mathrm d\tau\,\eta(\tau)\cos(\Omega\tau)\label{co11}\\
\gamma & :=\frac{1}{M\Omega}\int_0^\infty\mathrm d\tau\,\eta(\tau)\sin(\Omega\tau)\\
 D & :=\int_0^\infty\mathrm d\tau\,\nu(\tau)\cos(\Omega\tau)\\
 f & :=-\frac{1}{M\Omega}\int_0^\infty\mathrm d\tau\,\nu(\tau)\sin(\Omega\tau).\label{co14}
\end{align}
It is clear that the coupling of the oscillators between each other leads at the level of the master equation to double commutators containing operators that belong to different oscillators. Furthermore, the coefficients \eqref{coef1}-\eqref{coef4} for the coupled oscillators are no more Fourier transforms of $\nu_\alpha(\tau)$, $\eta_\alpha(\tau)$ with respect to the bare frequencies $\Omega_\alpha$ but with respect to the {\it eigenfrequencies} of the matrix $\mathbf{G}$. In the limit of vanishing coupling $g_{\alpha\beta}\rightarrow 0$ the matrix $\mathbf{G}$ becomes diagonal, which leads to $\tilde{\Omega}_{\alpha\beta}^2=D_{\alpha\beta}=\gamma_{\alpha\beta}=f_{\alpha\beta}=0$ for $\alpha\neq\beta$. As expected, one obtains $N$ independent decohering oscillators with coefficients \eqref{co11}-\eqref{co14}.\\
\indent In reality one is often confronted with a continuous distribution of frequencies and coupling constants of the environment. In this case one can introduce spectral densities $J_\alpha(\omega)$ (which together with the temperatures $T_\alpha$ describe all aspects of the reservoirs which are relevant for decoherence), so that \eqref{nk} and \eqref{dk} write
\begin{align}\nu_\alpha(\tau) & := \int_0^\infty\mathrm d\omega\,J_\alpha(\omega)\coth\left(\frac{\omega}{2T_\alpha}\right)\cos(\omega\tau)\\
\eta_\alpha(\tau) & := \int_0^\infty\mathrm d\omega\,J_\alpha(\omega)\sin(\omega\tau).
 \end{align}
At this point, a comparison with the master equation obtained by Cacheffo et al. in \cite{cac} is interesting. The authors of \cite{cac} consider the Hamiltonian \eqref{ham} for $\alpha=2$ oscillators but with additional position-momentum and momentum-momentum couplings and apply the Caldeira-Leggett approximation. If in \eqref{BM} one uses ohmic spectral densities (see \cite{sc1} for technical details and \cite{cal} for the physical meaning of ohmic dissipation)
\begin{equation}
\label{ohm}J_\alpha^{(ohm)}(\omega)=\frac{2}{\pi}M_\alpha\gamma_0^{(\alpha)}\frac{\omega}{1+(\omega/\Lambda_\alpha)^2}
 \end{equation}
with cut-off frequencies $\Lambda_\alpha$ and assumes that the typical frequency scale $\Omega=\max(\Omega_1,\Omega_2)$ of the system is lower than the frequency scale $\Lambda=\min(\Lambda_1,\Lambda_2)$ of the environment and that the temperatures $T_\alpha$ are high compared to the frequencies of the environment ($\Lambda_\alpha\ll T_\alpha$), one obtains the master equation 
\begin{equation}\label{CL}\begin{split}\frac{\mathrm d}{\mathrm d t}\rho_\mathcal{S}(t)= & -i[H_\mathcal{S},\rho_\mathcal{S}(t)]\\
& -\sum_{\alpha=1}^2  \Bigl(2M_\alpha\gamma_0^{(\alpha)}T_\alpha[X_\alpha,[X_\alpha,\rho_\mathcal{S}(t)]]+i\gamma_0^{(\alpha)}[X_\alpha,\{P_\alpha,\rho_\mathcal{S}(t)\}]\Bigr.\\
& \qquad-i M_\alpha\gamma_0^{(\alpha)}\Lambda_\alpha[X_\alpha,\{X_\alpha,\rho_\mathcal{S}(t)\}]\Bigr),\end{split}\end{equation}
derived in \cite{cac} for two linearly coupled oscillators, provided that the position-momentum and momentum-momentum couplings introduced therein are neglected. As the authors of \cite{cac} remark, the Caldeira-Leggett limit with only position-position coupling is trivial: it generates no new non-unitary contributions. It is interesting that the master equation (38) in \cite{cac} with (bilinear) position-momentum and momentum-momentum couplings is very similar to our master equation \eqref{BM} (of course the terms in $f_{\alpha\beta}$ are absent, as they vanish in the high temperature limit). Note that there is no problem to introduce position-momentum and momentum-momentum couplings in our model as well, this would merely change the matrix $\mathbf{G}$. As long as the Born-Markov approximation holds, we have therefore extended the master equation from \cite{cac} to arbitrary temperatures and spectral densities.\\
\indent To see the physical meaning of the coefficients \eqref{coef1}-\eqref{coef4}, we set up differential equations for the expectation values of position and momentum. This can be done using $\mathrm d/\mathrm d t \langle O\rangle_t=\tr(\mathrm d\rho_\mathcal{S}/\mathrm d t \,O)$ for time-independent observables $O$ and inserting \eqref{BM}. In this way one obtains the following Ehrenfest-like theorem:
\begin{equation}
\label{ehr}\frac{\mathrm d}{\mathrm d t}\begin{pmatrix}\langle \mathbf{X}\rangle_t\\ \langle \mathbf{P}\rangle_t\end{pmatrix}=\mathbf{A}_{\mathrm{diss}}\begin{pmatrix}\langle \mathbf{X}\rangle_t\\ \langle \mathbf{P}\rangle_t\end{pmatrix},
\end{equation}
where the position and momentum operators are contained in the vectors $\mathbf{X}:=(X_1,\dots,X_N)^T$ and $\mathbf{P}:=(P_1,\dots,P_N)^T$ and a dissipation matrix $\mathbf{A}_{\mathrm{diss}}$ is defined by
\begin{equation}
 \mathbf{A}_{\mathrm{diss}}:=\begin{pmatrix}\mathbf{0} & \mathbf{M}^{-1} \\ -\mathbf{g}' & -2\boldsymbol{\gamma}\end{pmatrix}
\end{equation}
with
\begin{align}
 \mathbf{M}:=\mathrm{diag}(M_1,\dots,M_N)\qquad\boldsymbol{\gamma}:=\begin{pmatrix}\gamma_{11} & \cdots & \gamma_{1N}\\ \vdots & & \vdots\\ \gamma_{N1} & \cdots & \gamma_{NN}\end{pmatrix}\\
\mathbf{g}':=\begin{pmatrix}M_1\Omega_1^{'\,2} & g'_{12} & \cdots & g'_{1N}\\ g'_{21} & \ddots & \ddots & \vdots\\ \vdots & \ddots & \ddots & g'_{N-1,N}\\ g'_{N1} & \cdots & g'_{N,N-1} & M_N\Omega_N^{'\,2}\end{pmatrix}.
\end{align}
Here we defined the renormalized coupling constants $g_{\alpha\beta}':=g_{\alpha\beta}+M_\beta\tilde{\Omega}_{\alpha\beta}^2$ (note that $g_{\alpha\beta}'\neq g_{\beta\alpha}'$ in general) and the renormalized frequencies $\Omega_\alpha'^2:=\Omega_\alpha^2+\tilde{\Omega}_{\alpha\alpha}^2$. The coefficients $\tilde{\Omega}_{\alpha\beta}^2$ therefore merely renormalize coupling constants and oscillator frequencies. From \eqref{ehr} one can see that $\gamma_{\alpha\alpha}$ describes the dissipation of the momentum of the  oscillator $\alpha$ and that $\gamma_{\alpha\beta}$ ($\alpha\neq\beta$) is responsible for the dissipation of the oscillator $\alpha$ caused by the coupling to the oscillator $\beta$. The formal solution of \eqref{ehr} is:
\begin{equation}
 \begin{pmatrix}\langle \mathbf{X}\rangle_t\\ \langle \mathbf{P}\rangle_t\end{pmatrix}=e^{\mathbf{A}_{\mathrm{diss}}t}\begin{pmatrix}\langle \mathbf{X}\rangle_{t=0}\\ \langle \mathbf{P}\rangle_{t=0}\end{pmatrix}
\end{equation}
\indent The differential equations for the second moments of $X_\alpha$ and $P_\alpha$ - in contrast to \eqref{ehr} - contain also the temperature dependent coefficients $D_{\alpha\beta}$ and $f_{\alpha\beta}$:
\begin{equation}
 \frac{\mathrm d }{\mathrm d t}\langle X_\alpha X_\beta\rangle_t=\frac{1}{2M_\beta}\langle\{X_\alpha,P_\beta\}\rangle_t+\frac{1}{2M_\alpha}\langle\{X_\beta,P_\alpha\}\rangle_t
\end{equation}
\begin{equation}
\begin{split}\frac{\mathrm d }{\mathrm d t}\langle P_\alpha P_\beta\rangle_t  = & D_{\alpha\beta}+D_{\beta\alpha}-2\sum_{\delta=1}^N\left(\gamma_{\alpha\delta}\langle P_\alpha P_\delta\rangle_t+\gamma_{\beta\delta}\langle P_\beta P_\delta\rangle_t\right)\\
& -\frac{1}{2}M_\beta\Omega_\beta'^2\langle\{X_\beta,P_\alpha\}\rangle_t-\frac{1}{2}M_\alpha\Omega_\alpha'^2\langle\{X_\alpha,P_\beta\}\rangle_t\\
& -\frac{1}{2}\left(\sum_{\delta=1,\,\delta\neq\beta}^N g_{\beta\delta}'\langle\{X_\delta,P_\alpha\}\rangle_t+\sum_{\delta=1,\,\delta\neq\alpha}^N g_{\alpha\delta}'\langle\{X_\delta,P_\beta\}\rangle_t\right)\end{split}
\end{equation}
\begin{equation}
 \begin{split}\frac{\mathrm d }{\mathrm d t}\langle \{X_\alpha, P_\beta\}\rangle_t= & -2f_{\beta\alpha}-2\sum_{\delta=1}^N\gamma_{\beta\delta}\langle\{X_\alpha,P_\delta\}\rangle_t+\frac{2}{M_\alpha}\langle P_\alpha P_\beta\rangle_t\\
& -2M_\beta\Omega_\beta'^2\langle X_\alpha X_\beta\rangle_t-2\sum_{\delta=1,\,\delta\neq\beta}^N g_{\beta\delta}'\langle X_\delta X_\alpha\rangle_t\end{split}
\end{equation}
As in the case of Quantum Brownian Motion \cite{sc1}, the coefficients $D_{\alpha\alpha}$ describe diffusion in momentum and decoherence in position basis of the oscillator $\alpha$. Therefore, the ``off-diagonal'' coefficients $D_{\alpha\beta}$ account for the decoherence of the oscillator $\alpha$ caused by the coupling to the oscillator $\beta$. A similar interpretation can be attributed to the anomalous diffusion coefficients $f_{\alpha\beta}$.

\subsection{Solution of Born-Markov Equation}

In contrast to the exact master equation, the Born-Markov equation \eqref{BM} can be solved easily. To do this, it is useful to change to a more suitable basis via the transformation \cite{unr,joo}
\begin{equation}
 \label{trans}\rho(\mathbf{k},\mathbf{\Delta},t):=\tr\left(\exp\left(i(\mathbf{k}\cdot\mathbf{X}+\mathbf{\Delta}\cdot\mathbf{P})\right)\rho_\mathcal{S}(t)\right)=\tr\left(e^{i\mathbf{k}\cdot\mathbf{\Delta}/2}e^{i\mathbf{k}\cdot\mathbf{X}}e^{i\mathbf{\Delta}\cdot\mathbf{P}}\rho_\mathcal{S}(t)\right),
\end{equation}
where $\mathbf{k}:=(k_1,\dots,k_N)$, $\mathbf{\Delta}:=(\Delta_1,\dots,\Delta_N)$. Physically, the coordinate $\mathbf{\Delta}$ measures the distance from the diagonal $\mathbf{x}=\mathbf{x}'$ in position basis, while $\mathbf{k}$ can be associated with the position \emph{on} the diagonal. Note also the formal similarity to the Weyl transform \cite{moy}. By applying \eqref{trans} on \eqref{BM} one obtains the following first order linear partial differential equation:
\begin{equation}\label{red-BM}
\begin{split}
  & \frac{\partial}{\partial t} \rho(\mathbf{k},\mathbf{\Delta},t) \\
 & =\Biggl[\sum_{\alpha=1}^N\left(\frac{k_\alpha}{M_\alpha}\frac{\partial}{\partial\Delta_\alpha}-M_\alpha\Omega_\alpha^{'\,2}\Delta_\alpha\frac{\partial}{\partial k_\alpha}-D_{\alpha\alpha}\Delta_\alpha^2-2\gamma_{\alpha\alpha}\Delta_\alpha\frac{\partial}{\partial\Delta_\alpha}+f_{\alpha\alpha}\Delta_\alpha k_\alpha\right)\Biggr.\\
 & \qquad\Biggl.+\sum_{\alpha\neq\beta}\left(-g'_{\alpha\beta}\Delta_\alpha\frac{\partial}{\partial k_\beta}-D_{\alpha\beta}\Delta_\alpha\Delta_\beta-2\gamma_{\alpha\beta}\Delta_\alpha\frac{\partial}{\partial\Delta_\beta}+f_{\alpha\beta}\Delta_\alpha k_\beta\right)\Biggr]\rho(\mathbf{k},\mathbf{\Delta},t)
\end{split}
\end{equation}
Equation \eqref{red-BM} can be solved using the method of characteristics. On a characteristic line $(\mathbf{k}(\tau),\mathbf{\Delta}(\tau),t(\tau))$ the original equation reduces to an ordinary differential equation:
\begin{equation}
 \label{BM-ch}\frac{\mathrm d}{\mathrm d\tau}\rho(\mathbf{k}(\tau),\mathbf{\Delta}(\tau),t(\tau))=(\mathbf{k}(\tau),\mathbf{\Delta}(\tau))\mathbf{B}_\mathrm{dec}(\mathbf{k}(\tau),\mathbf{\Delta}(\tau))^T,
\end{equation}
where the decoherence matrix $\mathbf{B}_\mathrm{dec}$ can be defined by
\begin{equation}
\mathbf{B}_\mathrm{dec}:=\begin{pmatrix}\mathbf{0} & \mathbf{0}\\-\mathbf{f} & \mathbf{D}\end{pmatrix}
\end{equation}
\begin{equation}
\mathbf{f}:=\begin{pmatrix}f_{11} & \cdots & f_{1N}\\ \vdots & & \vdots\\ f_{N1} & \cdots & f_{NN}\end{pmatrix}\qquad\mathbf{D}:=\begin{pmatrix}D_{11} & \cdots & D_{1N}\\ \vdots & & \vdots\\ D_{N1} & \cdots & D_{NN}\end{pmatrix}
\end{equation}
By comparing \eqref{red-BM} with $\frac{\mathrm d\rho}{\mathrm d\tau}=\frac{\mathrm d\mathbf{k}}{\mathrm d\tau}\cdot\nabla_\mathbf{k}\rho+\frac{\mathrm d\mathbf{\Delta}}{\mathrm d\tau}\cdot\nabla_\mathbf{\Delta}\rho+\frac{\mathrm d t}{\mathrm d\tau}\frac{\partial\rho}{\partial t}$ one obtains:
\begin{equation}
 \label{char}\frac{\mathrm d t}{\mathrm d\tau}=-1\qquad
\frac{\mathrm d}{\mathrm d\tau}\begin{pmatrix}\mathbf{k}(\tau)\\\mathbf{\Delta}(\tau)\end{pmatrix}=\mathbf{A}_\mathrm{diss}^T\begin{pmatrix}\mathbf{k}(\tau)\\\mathbf{\Delta}(\tau)\end{pmatrix}
\end{equation}
Solving \eqref{char} and inserting in \eqref{BM-ch} yields:
\begin{equation}
\rho(\tau)=\rho(\mathbf{k}_0,\mathbf{\Delta}_0,0)\exp\left((\mathbf{k}_0,\mathbf{\Delta}_0)\int_0^\tau e^{\mathbf{A}_\mathrm{diss}\tau'}\mathbf{B}_\mathrm{dec}e^{\mathbf{A}_\mathrm{diss}^T\tau'}\mathrm d\tau'\,(\mathbf{k}_0,\mathbf{\Delta}_0)^T\right)
\end{equation}
with $\mathbf{k}(\tau=0)=:\mathbf{k}_0$, $\mathbf{\Delta}(\tau=0)=:\mathbf{\Delta}_0$, $t(\tau=0)=0$. After retransformation into the time domain this gives the compact form
\begin{equation}\label{sol}
\rho(\mathbf{k},\mathbf{\Delta},t)=\rho_0\left((\mathbf{k},\mathbf{\Delta})\,e^{\mathbf{A}_\mathrm{diss}t}\right)\exp\left(-(\mathbf{k},\mathbf{\Delta})\,\mathbf{C}(t)(\mathbf{k},\mathbf{\Delta})^T\right)
\end{equation}
with initial condition $\rho(\mathbf{k},\mathbf{\Delta},0)=:\rho_0(\mathbf{k},\mathbf{\Delta})$ and decoherence kernel
\begin{equation}
\mathbf{C}(t):=e^{\mathbf{A}_\mathrm{diss}t}\int_0^t e^{-\mathbf{A}_\mathrm{diss}\tau}\,\mathbf{B}_\mathrm{dec}\,e^{-\mathbf{A}_\mathrm{diss}^T\tau}\,\mathrm d\tau\,e^{\mathbf{A}_\mathrm{diss}^Tt}.
\end{equation}
The $(\mathbf{k},\mathbf{\Delta})$-representation is related to the position representation via
\begin{equation}\label{pos}
 \langle \mathbf{x}\vert \rho_\mathcal{S}(t)\vert \mathbf{x}'\rangle =\int\mathrm d^N k\,e^{-i\mathbf{k}\cdot\frac{\mathbf{x}+\mathbf{x}'}{2}}\rho(\mathbf{k},\mathbf{x}-\mathbf{x}',t)
\end{equation}
(with $\mathbf{x}:=(x_1,\dots,x_N)^T$, $\mathbf{x}':=(x_1',\dots,x_N')^T$), as can be verified by inserting in \eqref{trans} and evaluating the trace in the position basis.\\
\indent It is interesting from a mathematical point of view that the solution \eqref{sol} factorizes into a part depending on the initial condition and a Gaussian part describing decoherence and therefore independent of the initial condition (a similar result was found for Quantum Brownian Motion in \cite{ven}). It is also surprising that the same matrix $\mathbf{A}_\mathrm{diss}$ that describes the evolution of the expectation values $\langle\mathbf{X}\rangle$ and $\langle\mathbf{P}\rangle$ enters in the part of the solution depending on the initial condition. If one compares \eqref{sol} with the solution of the single oscillator problem, one finds that the two solutions are formally equivalent. The only difference is that in the case of the single oscillator, the dissipation matrix $\mathbf{A}_\mathrm{diss}$ and decoherence matrix $\mathbf{B}_\mathrm{dec}$ are defined by:
\begin{equation}
 \mathbf{A}_{\mathrm{diss}}^{(N=1)}:=\begin{pmatrix}0 & 1/M \\ -M\Omega'^2 & -2\gamma\end{pmatrix}\qquad \mathbf{B}_{\mathrm{dec}}^{(N=1)}:=\begin{pmatrix}0 & 0\\-f & D\end{pmatrix}
\end{equation}
As expected, in the case of the coupled oscillators mixed terms $k_\alpha k_\beta$, $k_\alpha \Delta_\beta$ and $\Delta_\alpha\Delta_\beta$ ($\alpha\neq\beta$) appear in the Gaussian. In order to explicitly calculate the time-dependent matrix $\mathbf{C}(t)$, one has to diagonalize $\mathbf{A}_\mathrm{diss}$. Interestingly, this can be done analytically only in the case of $N=2$ interacting oscillators (for higher $N$ the order of the characteristic polynomial of $\mathbf{A}_\mathrm{diss}$ will be greater than 4 and therefore cannot be solved analytically, if no additional symmetries are assumed).

\section{Common Reservoir}\label{s3}

\begin{figure}[ht]
\centering
 \includegraphics[width=4cm,angle=270]{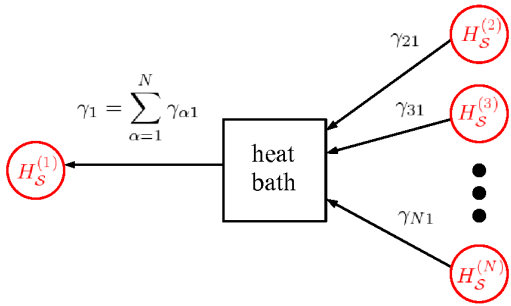}
\caption{Illustration of the influence of a common reservoir: The coupling to the same reservoir introduces reservoir-induced interactions, therefore e.g. the dissipation coefficient $\gamma_1$ of oscillator 1 is the sum of the dissipation coefficients $\gamma_{\alpha 1}$ for distinct reservoirs.}
\label{res}
\end{figure}

Now, we consider the case where the system described by the Hamiltonian $H_\mathcal{S}$ is coupled to a common reservoir
\begin{equation}
H_\mathcal{E}^{(c)}=\sum_{j=1}^{N_{HB}}\left(\frac{p_j^2}{2m_j}+\frac{1}{2}m_j \omega_j^2 x_j^2\right)
\end{equation}
via an interaction Hamiltonian
\begin{equation}
 H_I^{(c)}=\sum_{\alpha=1}^N X_\alpha\otimes\sum_{j=1}^{N_{HB}} c_j^{(\alpha)} x_j\equiv\sum_{\alpha=1}^N X_\alpha\otimes E_\alpha^{(c)}.
\end{equation}
The time evolution of the operator $X_\alpha$ in the interaction picture is therefore still given by \eqref{xev}, but the correlation function is now different. Using the annihilation and creation operators $b_j$, $b_j^\dagger$ ($j=1,\dots,N_{HB}$) of the heat bath, one obtains
\begin{equation}
 \mathcal{C}_{\alpha\beta}^{(c)}(\tau)=\sum_{j,l=1}^{N_{HB}}\frac{c_j^{(\alpha)} c_l^{(\beta)}}{2\sqrt{m_j\omega_jm_l\omega_l}}\left\langle\left(b_j^\dagger e^{i\omega_j\tau}+b_je^{-i\omega_j\tau}\right)\left(b_l+b_l^\dagger\right)\right\rangle_{\rho_\mathcal{E}}.
\end{equation}
Consequently
\begin{equation}
\label{coc}\mathcal{C}_{\alpha\beta}^{(c)}(\tau)=\nu^{(c)}_{\alpha\beta}(\tau)-i\eta^{(c)}_{\alpha\beta}(\tau)
\end{equation}
with noise kernel
\begin{equation}\nu^{(c)}_{\alpha\beta}(\tau):= \sum_{j=1}^{N_{HB}} \frac{c_j^{(\alpha)}c_j^{(\beta)}}{2m_j\omega_j}\coth\left(\frac{\omega_j}{2T}\right)\cos(\omega_j\tau)\equiv\int_0^\infty\mathrm d\omega\,J_{\alpha\beta}(\omega)\coth\left(\frac{\omega}{2T}\right)\cos(\omega\tau)
 \end{equation}
and dissipation kernel
\begin{equation}
 \eta^{(c)}_{\alpha\beta}(\tau):= \sum_{j=1}^{N_{HB}} \frac{c_j^{(\alpha)}c_j^{(\beta)}}{2m_j\omega_j}\sin(\omega_j\tau)\equiv\int_0^\infty\mathrm d\omega\,J_{\alpha\beta}(\omega)\sin(\omega\tau)
\end{equation}
(where $T$ is the temperature and $J_{\alpha\beta}(\omega)$ the spectral density of the heat bath). Comparing \eqref{coc} with \eqref{cod}, one sees that in the case of a common reservoir the correlation function is no longer diagonal in the indices $\alpha$ and $\beta$. In the following, we will only consider the case of equal coupling of the oscillators to the heat bath, i. e. $c_j^{(\alpha)}=c_j^{(\beta)}\equiv c_j$ and consequently $J_{\alpha\beta}(\omega)\equiv J(\omega)$ (for all $\alpha,\beta=1,\dots,N$). Using \eqref{coc}, one is therefore led to a different Born-Markov equation:
\begin{equation}\label{BM2}\begin{split}\frac{\mathrm d}{\mathrm d t}\rho_\mathcal{S}=-i[H_\mathcal{S},\rho_\mathcal{S}]-\sum_{\alpha,\beta=1}^N & \Bigl( D_{\beta} [X_\alpha,[X_\beta,\rho_\mathcal{S}]]+i\gamma_{\beta}[X_\alpha,\{P_\beta,\rho_\mathcal{S}\}]\Bigr.\\
& \Bigl.+f_{\beta} [X_\alpha,[P_\beta,\rho_\mathcal{S}]]+\frac{i}{2}M_\beta\tilde{\Omega}_{\beta}^2[X_\alpha,\{X_\beta,\rho_\mathcal{S}\}]\Bigr),\end{split}\end{equation}
where the coefficients are sums of the coefficients \eqref{coef1}-\eqref{coef4} for distinct reservoirs, i. e.
\begin{equation}
\label{sum}Y_\beta:=\sum_{\alpha=1}^N Y_{\alpha\beta}
\end{equation}
with $Y=\tilde{\Omega}^2,D,\gamma,f$, taking $T_\alpha\equiv T$, $J_\alpha(\omega)\equiv J(\omega)$ $\forall\alpha=1,\dots,N$ . Interestingly, the structure of the Born-Markov equation for a common reservoir is the same as for distinct reservoirs, only the coefficients, which contain information about the influence of the heat bath on the system, are different. Therefore, the density matrix for distinct reservoirs is formally equivalent to the one for a common reservoir. Physically, \eqref{sum} means that the oscillators are not only coupled by an explicit interaction but also indirectly via the interaction with the heat bath (figure \ref{res}). This feature of the common reservoir becomes particularly striking, if one considers the limit of vanishing coupling $g_{\alpha\beta}\rightarrow 0$. In this case, one obtains the master equation
\begin{equation}\begin{split}\frac{\partial}{\partial t}\rho_\mathcal{S}= -i[H_\mathcal{S},\rho_\mathcal{S}]-& \sum_{\alpha,\beta=1}^N \Bigl( D_{\beta}^{(g=0)} [X_\alpha,[X_\beta,\rho_\mathcal{S}]]+i\gamma_{\beta}^{(g=0)}[X_\alpha,\{P_\beta,\rho_\mathcal{S}\}]\Bigr.\\
& \Bigl.+f_{\beta}^{(g=0)} [X_\alpha,[P_\beta,\rho_\mathcal{S}]]+\frac{i}{2}M_\beta\left(\tilde{\Omega}_{\beta}^{(g=0)}\right)^2[X_\alpha,\{X_\beta,\rho_\mathcal{S}\}]\Bigr)\end{split}
\end{equation}
Despite the vanishing coupling constants $g_{\alpha\beta}$, the oscillators are mutually coupled by a reservoir-induced interaction, which is due to the fact that all oscillators are coupled to the {\it same} reservoir. In the next section, we will investigate this coupling mechanism in more detail.

\section{Case of Two Interacting Oscillators}\label{s4}

\subsection{Dependence of the Coefficients on the Interaction}

\begin{figure}
 \includegraphics[width=6cm]{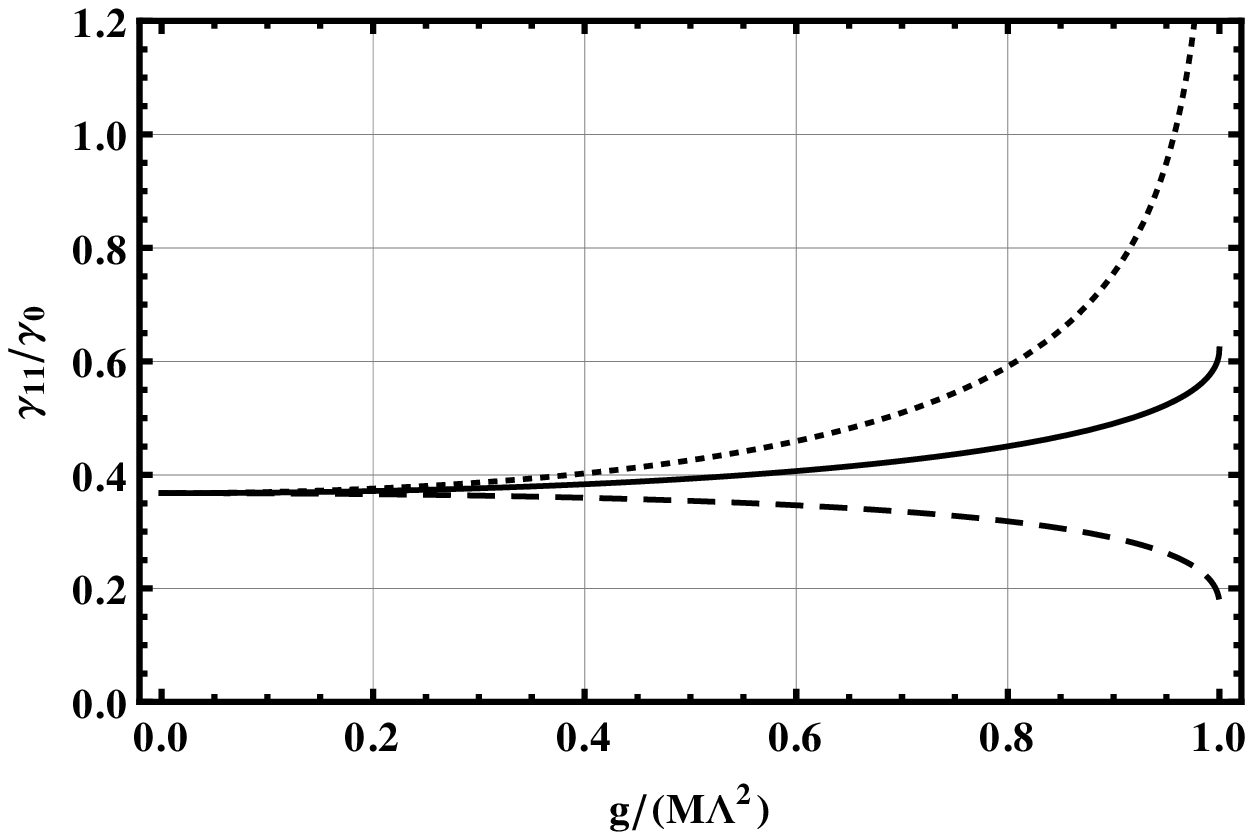}
\includegraphics[width=6cm]{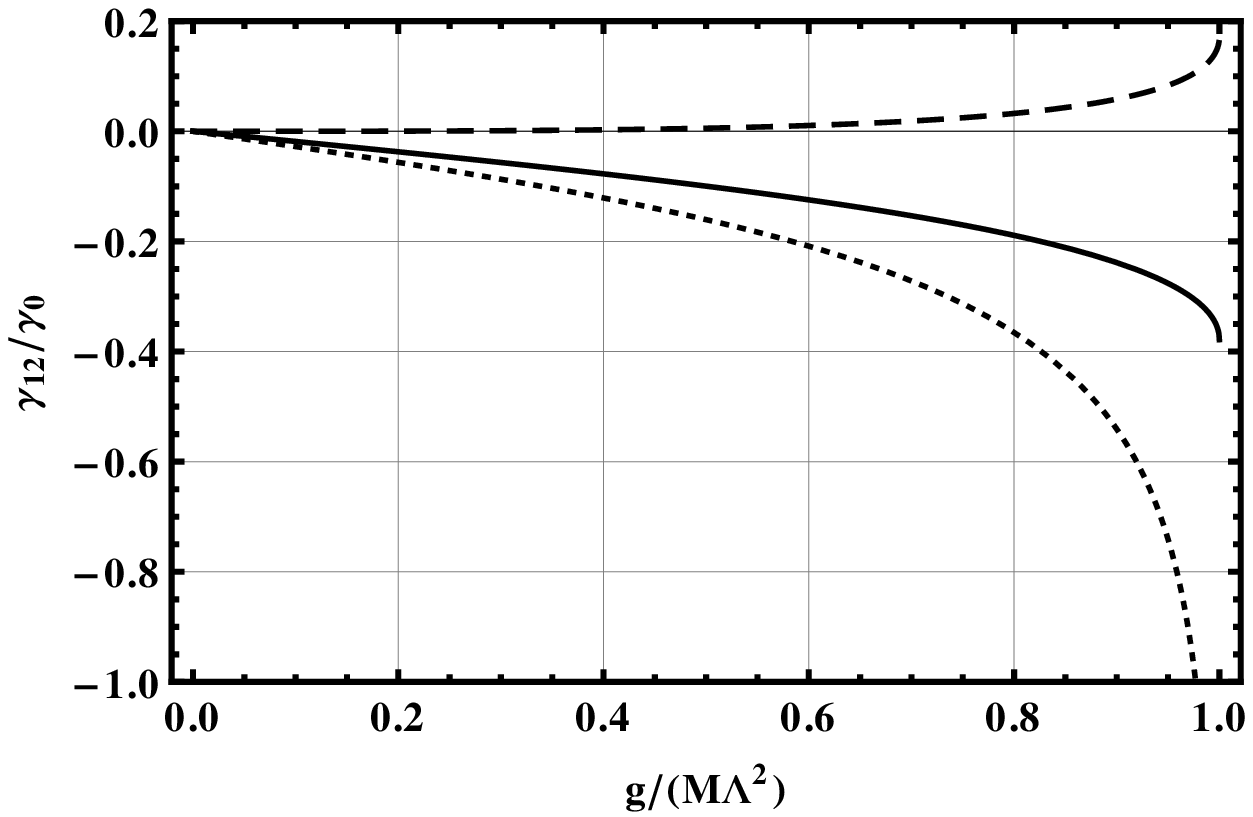}
\caption{Dependence of $\gamma_{11}$ (left) and $\gamma_{12}$ (right), normalized by $\gamma_0$, on the coupling strength $g$ (here: $r_+=1$, $r_-=0$). The dotted curves correspond to $s=0.5$ (sub-ohmic), the continuous ones to $s=1$ (ohmic) and the dashed ones to $s=2$ (super-ohmic). For subohmic dissipation $\gamma_{11}$ ($\gamma_{12}$) diverges strictly monotonically to $+\infty$ ($-\infty$), when $g\rightarrow g_\infty$. In the case of ohmic dissipation, one finds the same monotic behaviour, but no divergence occurs. For super-ohmic dissipation there are different types of curves, ranging from strictly monotonically increasing and decreasing curves to curves having a local minimum or maximum (here only the case of a strictly monotonically decreasing/increasing curve is depicted).}
\label{gam}
\end{figure}

\begin{figure}
\centering
 \includegraphics[width=6cm]{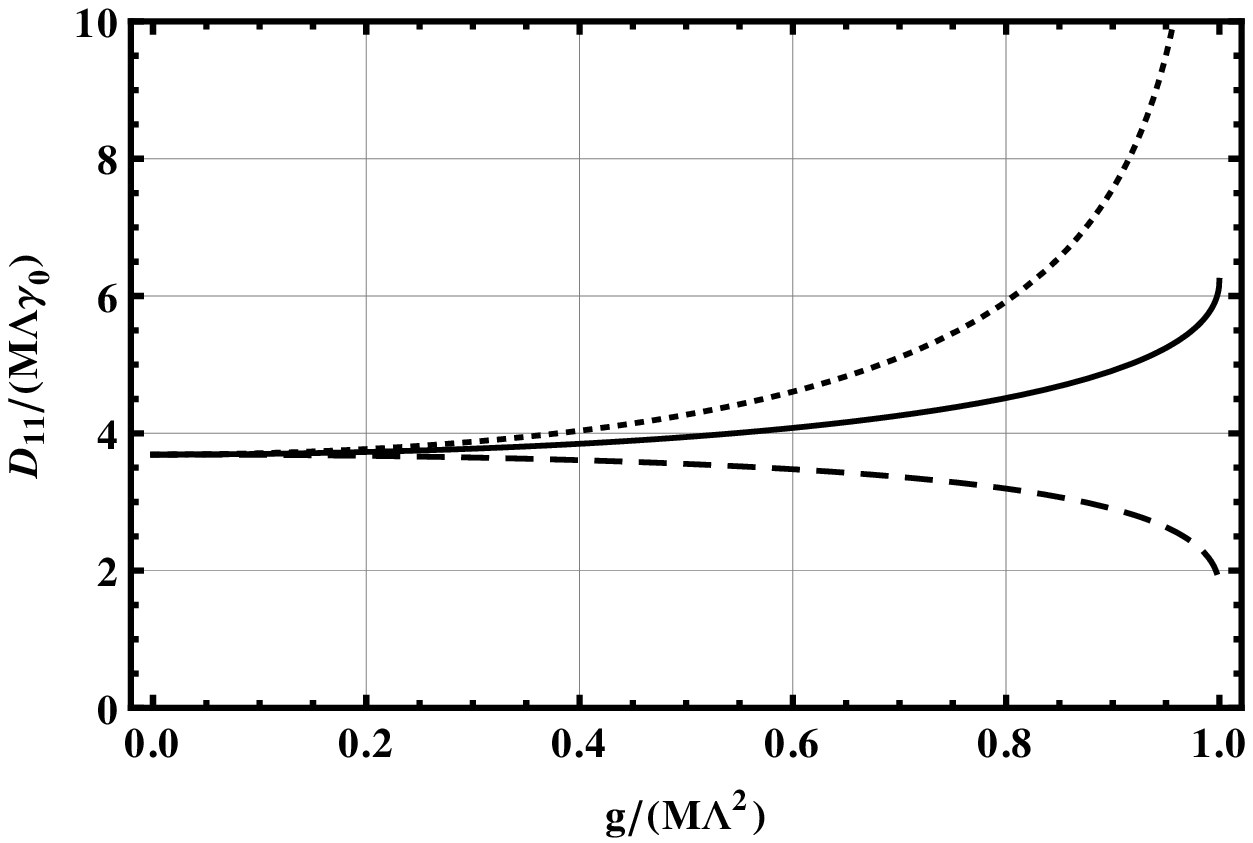}
\includegraphics[width=6cm]{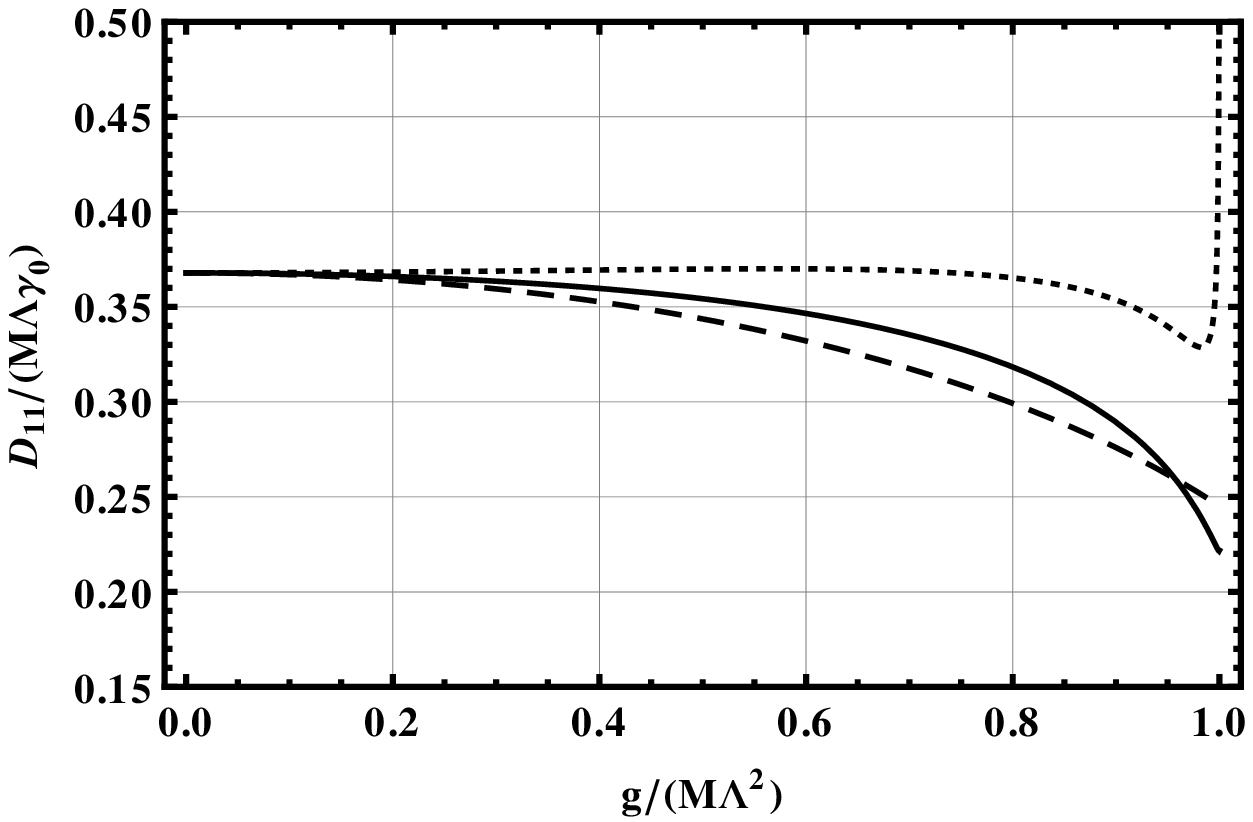}
\caption{Dependence of $D_{11}$, normalized by $\gamma_0M\Lambda$, on the coupling strength $g$ for $r_+=1$, $r_-=0$ and temperatures $T=5\Lambda$ (left) and $T=0.05\Lambda$ (right). The labelling of the curves is the same as in figure \ref{gam}. For high temperatures (left), compared with $\gamma_{\alpha\beta}$, the temperature $T$ simply rescales the $D_{\alpha\beta}$-axis. In the low temperature regime (right), the curves corresponding to sub-ohmic dissipation can present a minimum, while for ohmic dissipation the initially monotonically increasing curve is turned into a monotonically decreasing one. In the case of super-ohmic dissipation, the influence of temperature is smaller than for sub-ohmic and ohmic dissipation.}
\label{d}
\end{figure}

\begin{figure}
\centering
 \includegraphics[width=3.5cm,angle=270]{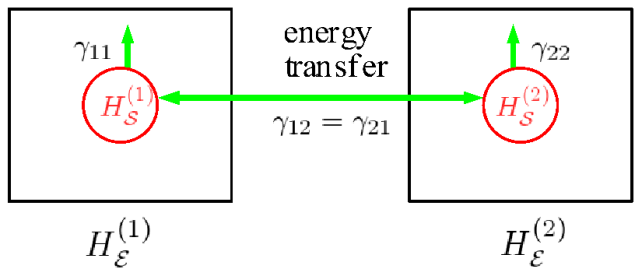}
\caption{Dissipation in the case of distinct reservoirs. The oscillators dissipate energy to their heat bath at the rates $\gamma_{11}$ and $\gamma_{22}$, respectively, while due to the coupling, energy is exchanged between the oscillators at equal rates $\gamma_{12}=\gamma_{21}$.}
\label{dis}
\end{figure}

\begin{figure}[ht]
\centering
 \includegraphics[width=6cm]{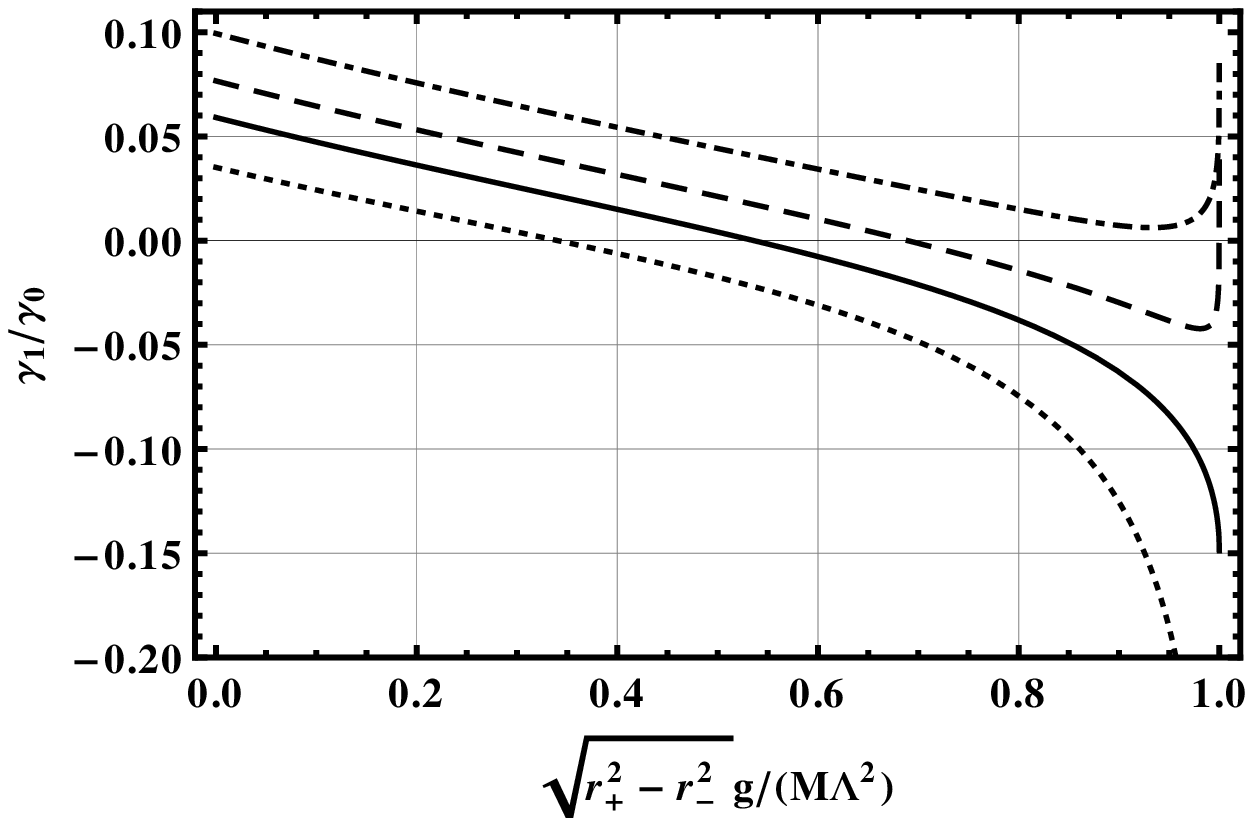}
 \includegraphics[width=6cm]{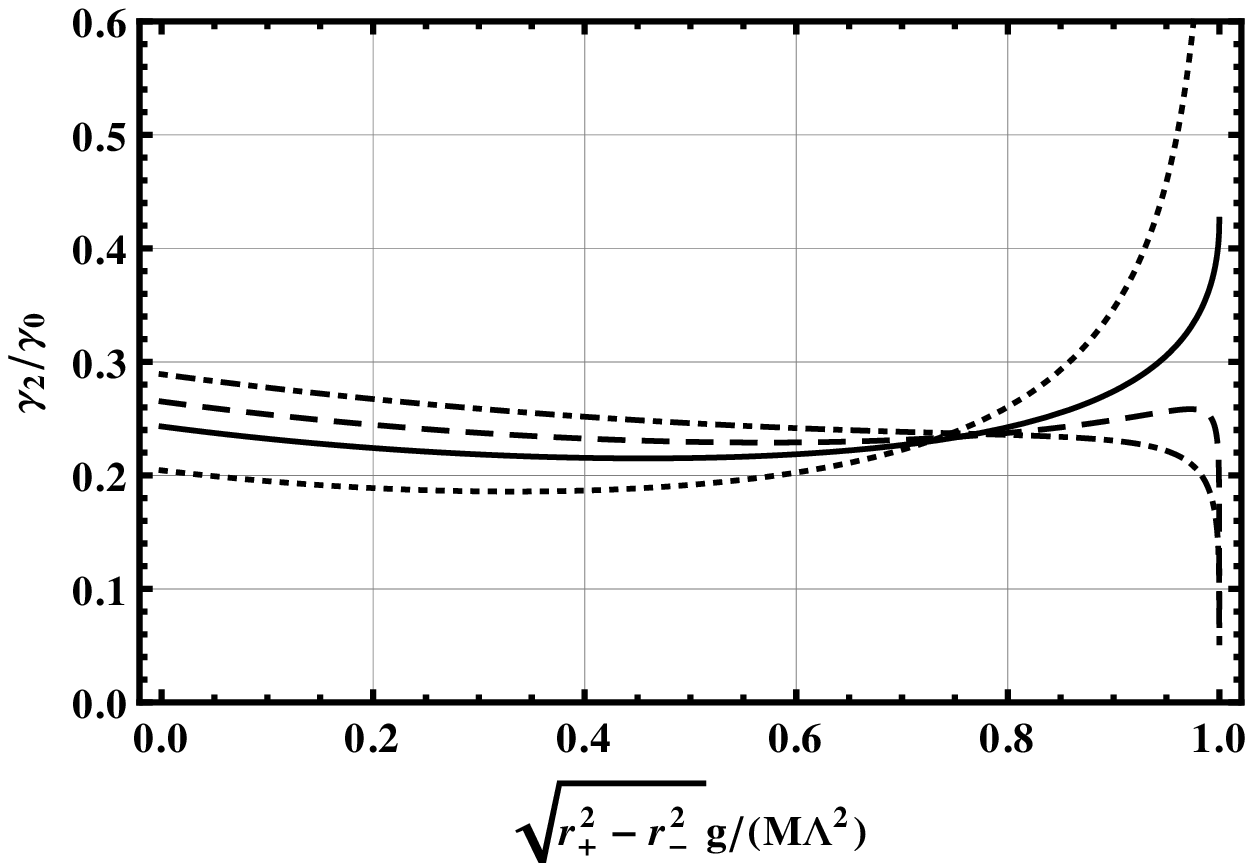}
\caption{Dissipation coefficients $\gamma_1$ (left) and $\gamma_2$ (right) for a common reservoir (here: $r_+=5$, $r_-=3$). The curves are plotted for different dissipation exponents $s$: dotted $s=0.5$, continuous $s=1$, dashed $s=1.25$, dotted-dashed $s=1.5$. For some values of $g$ the coefficient $\gamma_1$ can be negative, while $\gamma_2$ is always positive for $r_-\geq 0$.}
\label{ga1}
\end{figure}

\begin{figure}
\centering
 \includegraphics[width=4cm,angle=270]{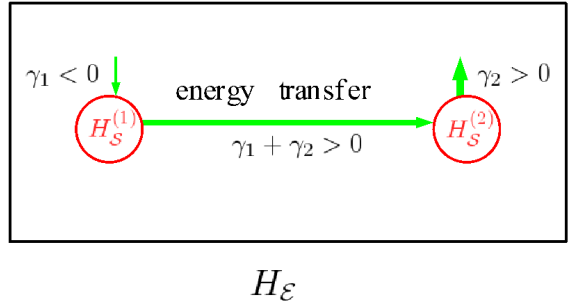}
\caption{Dissipation in the case of a common reservoir. For $r_->0$ and special values of $r_\pm$, $s$ and $\kappa$ it is $\gamma_1<0$, i. e. oscillator 1 gains energy from the common reservoir. Nevertheless, $\gamma_1+\gamma_2>0$ ensures that the other oscillator looses energy to the heat bath at a higher rate $\gamma_2>\lvert\gamma_1\rvert$. The coupling between the oscillators leads to the transfer of energy from oscillator 1 to oscillator 2, so that on the whole both oscillators are effectively subject to dissipation.}
\label{com}
\end{figure}

In this section, we will consider a specific physical situation described by the commonly used spectral density \cite{gra,hu,leg}
\begin{equation}
 \label{spe}J(\omega)=\frac{2}{\pi}M\gamma_0\omega^s\Lambda^{1-s} e^{-\omega/\Lambda},
\end{equation}
where $\Lambda$ is the cut-off frequency and $\gamma_0>0$ the coupling strength between the system and the reservoir. The exponent $s>0$ describes the type of dissipation: $0<s<1$ sub-ohmic, $s=1$ ohmic and $s>1$ super-ohmic dissipation. By varying the exponent $s$, one is a able to see the passage from local ($s=1$) to non-local dissipation ($s\neq 1$) and from white ($s=1$, high temperature) to coloured noise \cite{gra,hu}.\\
\indent If one wants to understand the physics of the system, the first step is to evaluate the coefficients appearing in the master equation. In the following, we will concentrate on the two-oscillator system with equal masses $M_1=M_2\equiv M$ and coupling constant $g_{12}=g_{21}\equiv g$ and will assume the same spectral density \eqref{spe} and temperature $T$ for both reservoirs. Furthermore, we will focus on the coefficients $\gamma_{\alpha\beta}$ and $D_{\alpha\beta}$, which are particularly easy to calculate, as they are double Fourier transforms of $J(\omega)$  and $J(\omega)\coth(\omega/(2T))$, respectively. Moreover, $\gamma_{\alpha\beta}$ and $D_{\alpha\beta}$ are expected to control the dynamics of decoherence, as it is the case with $\gamma$ and $D$ for Quantum Brownian Motion \cite{joo}. For distinct reservoirs one obtains:
\begin{equation}
\label{gad}\begin{split}\gamma_{11} = & \frac{\gamma_0}{2\sqrt{\kappa^2+r_-^2}}\\ & \times\left[\left(\sqrt{\kappa^2+r_-^2}+r_-\right)\left(r_++\sqrt{\kappa^2+r_-^2}\right)^{(s-1)/2}e^{-\sqrt{r_++\sqrt{\kappa^2+r_-^2}}}\right.\\
& \quad\left.+\left(\sqrt{\kappa^2+r_-^2}-r_-\right)\left(r_+-\sqrt{\kappa^2+r_-^2}\right)^{(s-1)/2}e^{-\sqrt{r_+-\sqrt{\kappa^2+r_-^2}}}\right]\end{split}
\end{equation}
\begin{equation}
 \label{gao}\begin{split}\gamma_{12}=\gamma_{21}=  \frac{\gamma_0\kappa}{2\sqrt{\kappa^2+r_-^2}} & \left[\left(r_++\sqrt{\kappa^2+r_-^2}\right)^{(s-1)/2}e^{-\sqrt{r_++\sqrt{\kappa^2+r_-^2}}}\right.\\
& \left.-\left(r_+-\sqrt{\kappa^2+r_-^2}\right)^{(s-1)/2}e^{-\sqrt{r_+-\sqrt{\kappa^2+r_-^2}}}\right]\end{split}
\end{equation}
\begin{equation}
\begin{split}D_{11}= & \frac{\gamma_0M\Lambda}{2\sqrt{\kappa^2+r_-^2}}\\
& \times\left[\left(\sqrt{\kappa^2+r_-^2}+r_-\right)\left(r_++\sqrt{\kappa^2+r_-^2}\right)^{s/2}e^{-\sqrt{r_++\sqrt{\kappa^2+r_-^2}}}\right.\\
& \qquad\qquad\qquad\qquad\qquad\qquad\qquad\times\coth\left(\frac{\sqrt{r_++\sqrt{\kappa^2+r_-^2}}}{2\tilde{T}}\right)\\
& \quad+\left(\sqrt{\kappa^2+r_-^2}-r_-\right)\left(r_+-\sqrt{\kappa^2+r_-^2}\right)^{s/2}e^{-\sqrt{r_+-\sqrt{\kappa^2+r_-^2}}}\\
&\left. \qquad\qquad\qquad\qquad\qquad\qquad\qquad\times\coth\left(\frac{\sqrt{r_+-\sqrt{\kappa^2+r_-^2}}}{2\tilde{T}}\right)\right]\end{split}
\end{equation}
\begin{equation}
 \begin{split}D_{12} & =D_{21} =\frac{\gamma_0\kappa M\Lambda}{2\sqrt{\kappa^2+r_-^2}}\\
& \times\left[\left(r_++\sqrt{\kappa^2+r_-^2}\right)^{s/2}e^{-\sqrt{r_++\sqrt{\kappa^2+r_-^2}}}\coth\left(\frac{\sqrt{r_++\sqrt{\kappa^2+r_-^2}}}{2\tilde{T}}\right)\right.\\
& \left.\quad-\left(r_+-\sqrt{\kappa^2+r_-^2}\right)^{s/2}e^{-\sqrt{r_+-\sqrt{\kappa^2+r_-^2}}}\coth\left(\frac{\sqrt{r_+-\sqrt{\kappa^2+r_-^2}}}{2\tilde{T}}\right)\right],\end{split}
\end{equation}
where $r_\pm:=(\Omega_1^2\pm\Omega_2^2)/(2\Lambda^2)$, $\kappa:=g/(M\Lambda^2)$, $\tilde{T}:=T/\Lambda$. The coefficients $\gamma_{22}$ and $D_{22}$ can be obtained from $\gamma_{11}$ and $D_{11}$ by making the substitution $r_-\rightarrow -r_-$. Before starting to analyze $\gamma_{\alpha\beta}$ and $D_{\alpha\beta}$, it should be said that all coefficients in the master equation are only defined for $g^2<M^2\Omega_1^2 \Omega_2^2=:g_\infty^2$. This is an anomality due to the specific choice of the coupling between the oscillators, which occurs when one diagonalizes $H_\mathcal{S}$.\\
\indent If one takes a closer look at the dissipation coefficients $\gamma_{\alpha\beta}$ (figure \ref{gam}), one sees that for sub-ohmic dissipation they are strictly monotonic with $g$ and diverge in the limit $g\rightarrow g_\infty$ (to $+\infty$ for $\alpha=\beta$ and to $-\infty$ for $\alpha\neq\beta$). For ohmic dissipation $\gamma_{\alpha\beta}$ is also strictly monotic with $g$ (increasing for $\alpha=\beta$ and decreasing for $\alpha\neq\beta$) but does not diverge. It is also interesting that for $\alpha\neq\beta$ the dissipation coefficients are always negative or zero in the case of ohmic and sub-ohmic dissipation. On the contrary, for $\alpha=\beta$ the coefficients are always positive, independently of $s$. In the super-ohmic case, the form of $\gamma_{\alpha\beta}$ as a function of $g$ explicitly depends on $r_\pm$ and $\tilde{T}$. In particular, $\gamma_{\alpha\beta}(g)$ can present a minimum or maximum for values of $s$ near 1.\\
Concerning the decoherence coefficients $D_{\alpha\beta}$, one finds that for high temperatures $T\gg\max(\Omega_1,\Omega_2)$ the relation
\begin{equation}\label{htp}
 D_{\alpha\beta}\approx 2MT\gamma_{\alpha\beta}
\end{equation}
holds. This relation is well known from the Caldeira-Leggett equation \cite{cal}, but in contrast to this, \eqref{htp} not only holds for ohmic but also for sub- and superohmic spectral densities. Moreover, the above result is completely independent of the cut-off frequency $\Lambda$ of the environment, which is not the case for the Caldeira Leggett approximation (see section \ref{ibm}). Consequently, for high temperatures, $D_{\alpha\beta}$ depends in exactly the same manner on the interaction strength $g$ as $\gamma_{\alpha\beta}$, the temperature simply rescales the $D_{\alpha\beta}$-axis (see figure \ref{d}, only $D_{11}$ is depicted). For low temperatures there exist corrections to \eqref{htp}: In the case of sub-ohmic dissipation, $D_{\alpha\alpha}$ ($D_{\alpha\beta}$, $\alpha\neq\beta$) still diverges to $+\infty$ ($-\infty$), but now a local minimum (maximum) can occur. For ohmic dissipation, decreasing temperature turns the initially stricly monotonically increasing (decreasing) curve of $D_{\alpha\alpha}$ ($D_{\alpha\beta}$) into a strictly monotonically decreasing (increasing) one. For super-ohmic dissipation the situation is more complex, but in general the temperatures necessary to cause deviations from the high temperature behaviour are lower than those for subohmic and ohmic dissipation. Independently of the temperature, $D_{\alpha\alpha}$ is always positive, while this need not to be the case for $D_{\alpha\beta}$ ($\alpha\neq\beta$). As one sees, the coupling of the oscillators leads to a rich behaviour already at the level of the coefficients $\gamma_{\alpha\beta}$ and $D_{\alpha\beta}$. This richness arises from the effect of the high frequency modes of the environment on the system; on the contrary, $\gamma_{\alpha\beta}$ and $D_{\alpha\beta}$ are \emph{independent} of the interaction strength $g$ in the Caldeira-Leggett limit (which corresponds to a long wavelength approximation, see \eqref{CL}).\\
\indent Let us now discuss the physical implications of the above mathematical observations. Because of the formal similarity of $\gamma_{\alpha\beta}$ and $D_{\alpha\beta}$ we will limit ourselves to the dissipation coefficients $\gamma_{\alpha\beta}$. From a quantum-optical point of view, the fact that $\gamma_{\alpha\beta}$ and $D_{\alpha\beta}$ are double Fourier transforms allows us to interpret them as a resonance of the environmental eigenmodes at the eigenfrequencies $\sqrt{r_+\pm\sqrt{\kappa^2+r_-^2}}$ of the system. As the eigenfrequency $\sqrt{r_+-\sqrt{\kappa^2+r_-^2}}$ is going to zero for $g\rightarrow g_\infty$, it is supposed that the divergency for $g\rightarrow g_\infty$ in the sub-ohmic case is connected with the divergence $J'(\omega\rightarrow 0)=+\infty$. If one keeps in mind that $\gamma_{\alpha\alpha}$ discribes the dissipation to the reservoir of the oscillator $\alpha$, it is clear that $\gamma_{\alpha\alpha}$ is always positive, for energy is transferred to the heat bath until the oscillator is slowed down to zero. That $\gamma_{\alpha\beta}$ ($\alpha\neq\beta$), which describes the dissipation of oscillator $\alpha$ due to the coupling to oscillator $\beta$, can become non-positive is not a problem, since in the two-oscillator system $\gamma_{\alpha\beta}=\gamma_{\beta\alpha}$ and an equal amount of energy is exchanged between the oscillators. This situation is illustrated in figure \ref{dis}.\\
\indent When both oscillators are coupled to the same heat bath, the situation is different: As $\gamma_{12}=\gamma_{21}$ can be negative, it is possible that $\gamma_1=\gamma_{11}+\gamma_{21}$ and $\gamma_2=\gamma_{22}+\gamma_{12}$ can also become negative. Adding \eqref{gad} and \eqref{gao}, one finds out that the possibility of $\gamma_1$ and $\gamma_2$ to become negative depends on the sign of $r_-$. More precisely, the only term in the expression for $\gamma_1$ that can be negative is $\sqrt{\kappa^2+r_-^2}-\kappa-r_-$, which is only the case for $r_->0$ (corresponding to $\Omega_1>\Omega_2$). In the same manner, the only negative term in $\gamma_2$ is $\sqrt{\kappa^2+r_-^2}-\kappa+r_-$ for $r_-<0$ ($\Omega_1<\Omega_2$). For $r_-=0$ ($\Omega_1=\Omega_2$) the coefficients $\gamma_1$ and $\gamma_2$ are always positive. Hence, $\gamma_1$ and $\gamma_2$ cannot become \emph{both} non-positive. Moreover, $\gamma_1+\gamma_2$ is always positive, which ensures that $\gamma_2>\lvert\gamma_1\rvert$ (for $\gamma_1<0$) and $\gamma_1>\lvert\gamma_2\rvert$ (for $\gamma_2<0$) are hold, i. e. energy is lost to the reservoir at a higher rate than it is gained. Under special circumstances (that depend in a complex way on the parameters $s$, $\kappa$ and $r_\pm$) $\gamma_1$ (or $\gamma_2$) is indeed negative (figure \ref{ga1}). Let us assume that $\Omega_1>\Omega_2$ and $\gamma_1<0$. In this case, oscillator 1 gains energy from the common heat bath at the rate $\lvert\gamma_1\rvert$ (figure \ref{com}). At the same time, oscillator 2 looses energy to the heat bath at a bigger rate $\gamma_2>\lvert\gamma_1\rvert$. Due to the coupling of the oscillators, an excess of energy is transferred from oscillator 1 to oscillator 2. This unexpected mechanism leads to the effective dissipation of {\it both} oscillators and is only possible because the oscillators are coupled to the {\it same} reservoir. Evidently, such a situation cannot be found by using the Caldeira-Leggett approximation, for which the coefficients $\gamma_{\alpha\beta}$ are always positive or zero. The same arguments hold in the case of the decoherence coefficients $D_{\alpha\beta}$. 
\newpage

\subsection{Reproduction of the Classical Motion and Quantum Corrections}

\begin{figure}
\centering
 \includegraphics[width=6cm]{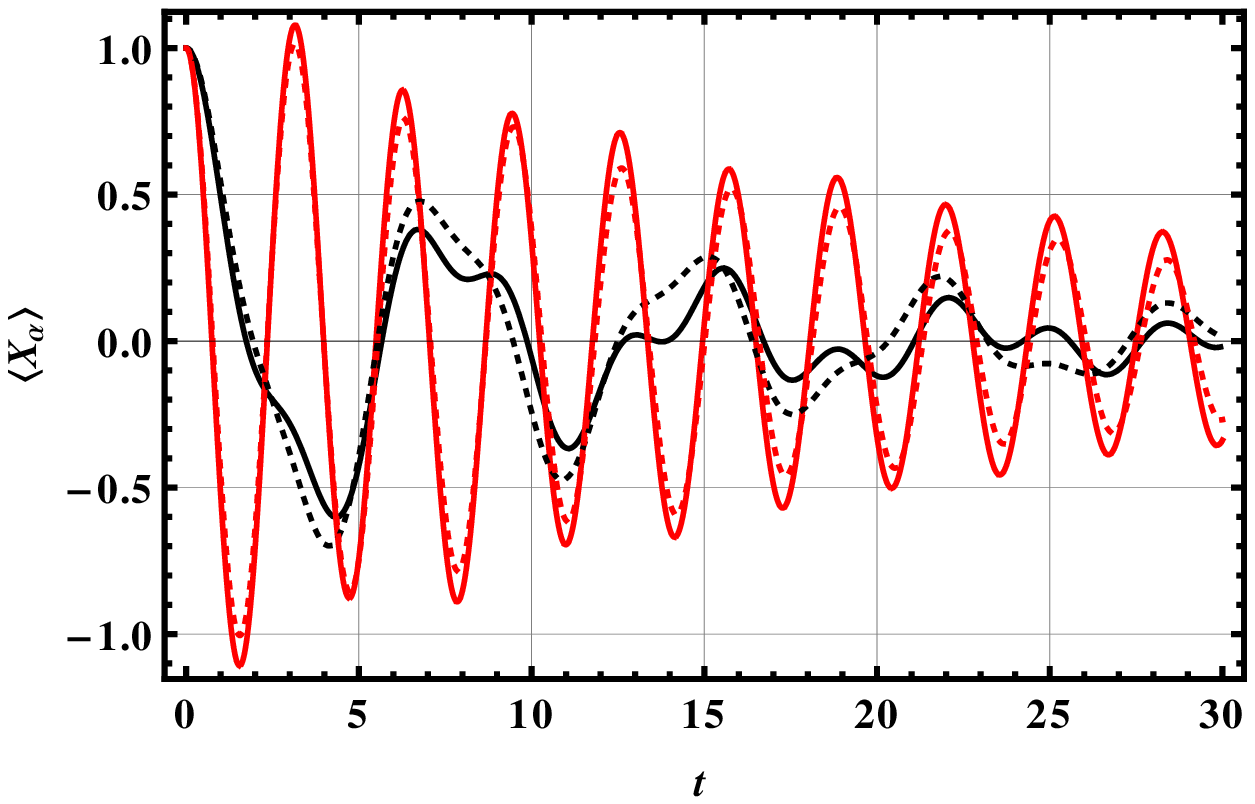}
 \includegraphics[width=6cm]{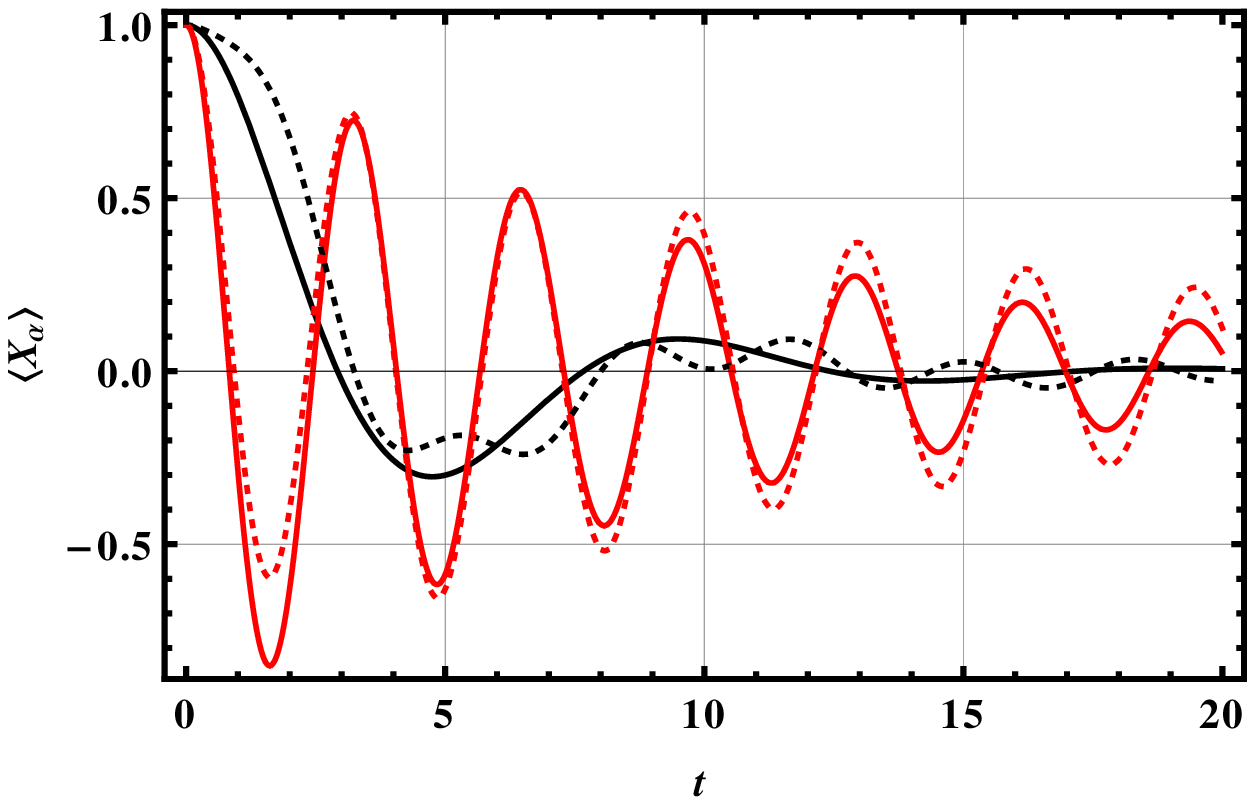}
\caption{Temporal evolution of the expectation values of position for ohmic dissipation ($\Lambda=1$, $M=1$, $\Omega_1=1$, $\Omega_2=2$). The initial conditions are $\langle X_1\rangle_{t=0}=\langle X_2\rangle_{t=0}=1$, $\langle P_1\rangle_{t=0}=\langle P_2\rangle_{t=0}=0$. The motion of oscillator 1 is depicted in black, the one of oscillator 2 in red; continuous curves represent distinct, dashed ones common reservoirs. On the left: two coupled oscillators ($g=0.5$, $\gamma_0=0.2$) which show the classically expected damped oscillations. The difference between distinct and common reservoirs is most striking for vanishing coupling $g=0$ (on the right, $\gamma_0=0.5$): For distinct reservoirs one obtains two independent damped oscillations, whereas in the case of a common reservoir the two oscillations are coupled by a reservoir-induced interaction.}
\label{cla}
\end{figure}

In this subsection, we will solve the differential equation system \eqref{ehr} for the expectation values of position for ohmic dissipation. This can be done analytically by using standard methods, but the resulting expressions are lengthy. We have therefore refrained from stating them explicitly. Instead of the above exponential cut-off we will use a Lorentz cut-off function as in \eqref{ohm}, which has the advantage that there exist closed expressions also for $\tilde{\Omega}_{\alpha\beta}^2$ and $f_{\alpha\beta}$ (only for the case $T=\Lambda/(2\pi)$) \cite{sc1} and not only for $\gamma_{\alpha\beta}$ and $D_{\alpha\beta}$. Note that the results of subsection 4.1 for $s=1$ can also be obtained by using the spectral density \eqref{ohm}. The choice of the cut-off function is therefore physically irrelevant. If one plots the solutions of \eqref{ehr}, one obtains the classically expected result: two coupled damped oscillations (figure \ref{cla}). In this example, due to the coupling, the oscillator with the lower frequency will follow the motion of the oscillator with the higher frequency. Figure \ref{cla} shows that the evolution of the mean position is not the same for distinct and common reservoirs, e. g. the low frequency oscillation takes longer to follow the high frequency oscillation in the case of a common reservoir. The differences between the two heat bath models are the bigger, the greater the parameters $\gamma_0$, $\Lambda$ and $g$ are. As already mentioned in section \ref{s3}, the case of vanishing coupling $g=0$ is particularly interesting: For distinct reservoirs, one obtains two independent damped oscillations, whereas in the case of a common reservoir a reservoir-induced interaction couples the two oscillations (figure \ref{cla}). Before concluding this section, it has to be said that \eqref{ehr} alone, with arbitrary coefficients $\tilde{\Omega}_{\alpha\beta}^2$ and $\gamma_{\alpha\beta}$, also admits a bunch of unphysical solutions (e. g. exponentially increasing oscillations). Only a physically reasonable spectral density, which enters in the quantum mechanical calculation of $\tilde{\Omega}_{\alpha\beta}^2$ and $\gamma_{\alpha\beta}$, ensures physical solutions (for the calculation of the spectral density from microscopic models see \cite{leg}).

\begin{figure}
\centering
 \includegraphics[width=6cm]{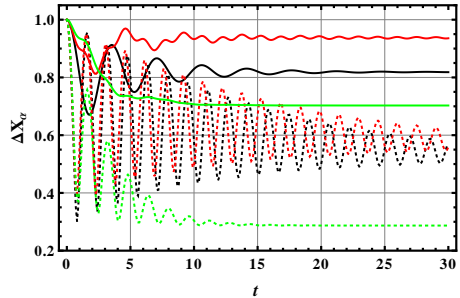}
\caption{Temporal evolution of the uncertainties of two coupled oscillators in separated reservoirs for ohmic dissipation ($\gamma=0.2$, $\Lambda=1$, $T=1/(2\pi)$, $M=1$, $\Omega_1=1$, $\Omega_2=2$). The initial uncertainties are chosen to be $\Delta X_1(0)=\Delta X_2(0)=1$. Continuous curves represent $\Delta X_1$, dashed ones $\Delta X_2$. The saturation values are not the same for uncoupled (black) and coupled (red, $g=1$) oscillators. For small times $t$ the uncertainties calculated by using the Caldeira-Leggett approximation (green, $g=1$, see \eqref{CL}) differ not much from those calculated with the help of the Born-Markov approximation. For larger times the differences are considerable.}
\label{dx}
\end{figure}

\indent To conclude this section, we briefly discuss how the interaction between the oscillators influences the \emph{quantum mechanical} properties of the system. For this purpose, we calculate the quantum mechanical uncertainties $\Delta X_\alpha(t)=\langle X_\alpha^2\rangle_t-\langle X_\alpha\rangle_t^2$ from the reduced density matrix for two initial Gaussian wave packets, described by the density matrix
\begin{equation}
 \rho_\mathcal{S}(\mathbf{x},\mathbf{x}',0)=\prod_{\alpha=1}^2\frac{1}{\sqrt{2\pi\Delta X_\alpha(0)^2}}\exp\left(-\frac{x_\alpha^2+x_\alpha^{'\,2}}{4\Delta X_\alpha(0)^2}\right).
\end{equation}
As in the case of one harmonic oscillator \cite{ven}, the uncertainties saturate after some time. But unlike the expectation values $\langle X\rangle_t$, which end in the same final state of vanishing amplitude, no matter how strong the interaction between the oscillators is, the uncertainties reach different values for $t\rightarrow\infty$, depending on the interaction strength (figure \ref{dx}, these values do not depend on the initial conditions). This means that \emph{quantum mechanically} an effect of the interaction remains even at $t\rightarrow\infty$. In figure \ref{dx}, one also sees that for low temperatures the uncertainties obtained by the Caldeira-Leggett approximation differ considerably from those calculated by using the Born-Markov approximation, especially for large times. It is also worth noting that for the studied temperature $T=\Lambda/(2\pi)$ the Caldeira-Leggett uncertainties systematically underestimate the Born-Markov uncertainties and relax more rapidly into their final values.

\section{Summary and Outlook}

In the present paper we worked out how to calculate the density matrix for a system of linearly coupled oscillators in a bosonic heat bath using the Born-Markov approximation. By having done this, we complemented previous results from \cite{cac}, which are only valid in the Caldeira-Leggett limit. As long as the Born-Markov approximation holds, we are now able to access the non-ohmic, low temperature regime. In the case of the two-oscillator system we found that the expectation values of position reproduce the classically expected result of coupled damped oscillations with certain renormalized coefficients. As for the quantum uncertainties, they are modified by the coupling even in the limit $t\rightarrow\infty$. Moreover, we have explictly seen that the coupling to a common reservoir leads to reservoir-induced interactions. In this context, using a specific spectral density, we found an interesting dissipation mechanism for a common reservoir, which is based on the fact that energy can be tranferred via the heat bath. This mechanism is due to the influence of the high frequency modes of the environment on the coupled system and is therefore not observable in the Caldeira-Leggett limit. The above results should encourage further studies in decoherence of interacting systems, in particular regarding the coupling to a common reservoir \cite{po1,cac}. We note that the presence of several reservoirs leads to interesting effects already for classical particles \cite{ono}. The calculation of other interesting physical quantities, like the entanglement of the subsystems, is postponed to a later work. In addition, it would also be interesting to study the effects that occur in the strong coupling, non-Markovian regime.

\section*{Acknowledgements} 
I would like to thank Haye Hinrichsen for supporting this work and Michelle Natera Comte for fruitful discussions.

\end{document}